\documentclass[showpacs,a4paper,nofootinbib]{revtex4}

\usepackage{amsmath,amssymb,amsthm}
\usepackage{graphicx}% Include figure files
\usepackage{psfrag}% use latex text in ps figures
\usepackage{bm}% bold math

\newcommand{\ds}[0]{d_{\rm s}}
\newcommand{\dc}[0]{d_{\rm c}}
\newcommand{\dcb}[0]{{\bar d}_{\rm c}}
\newcommand{\dsb}[0]{{\bar d}_{\rm s}}
\newcommand{\cs}{c_*}
\newcommand{\bb}{\bar b}
\newcommand{\tf}{\tilde f}
\newcommand{\ftz}{\widehat{f}_z}

\newcommand{\G}[0]{{G}}
\newcommand{\Gens}[0]{{\cal G}}
\newcommand{\ghat}[0]{\widehat{g}}
\newcommand{\Pt}[0]{\widehat{P}}
\newcommand{\Gt}[0]{\widehat{G}}
\newcommand{\Et}[3]{{\widehat E}^{#1}_{#2}[{#3}]}
\newcommand{\Etl}[1]{\Et{\lam}{}{#1}}
\newcommand{\Etv}[1]{\Et{V}{}{#1}}
\newcommand{\Etlr}[1]{\Et{\lam}{r}{#1}}
\newcommand{\Etvr}[1]{\Et{V}{r}{#1}}

\newcommand{\Pb}[0]{P}
\newcommand{\Gb}[0]{G}
\newcommand{\Eb}[3]{E^{#1}_{#2}[{#3}]}
\newcommand{\Ebl}[1]{\Eb{\lam}{}{#1}}
\newcommand{\Ebv}[2][V]{\Eb{#1}{}{#2}}
\newcommand{\Eblr}[1]{\Eb{\lam}{r}{#1}}
\newcommand{\Ebvr}[1]{\Eb{V}{r}{#1}}
\newcommand{\Eblm}[1]{\Eb{\lam}{r-1}{#1}}

\newcommand{\V}[1]{V_{#1}}
\newcommand{\VV}[2]{V_{#1,#2}}
\newcommand{\cF}[0]{{\cal F}}
\newcommand{\expv}[1]{\langle {#1} \rangle}
\newcommand{\gexpv}[1]{\langle\!\langle {#1} \rangle\!\rangle}
\newcommand{\avg}[1]{\overline{#1}}  
\newcommand{\sch}[2]{\!\!\sum_{\substack{#2 \text{ branches} \\ \text{of }#1}}}

\newcommand{\bt}{\beta}

\newcommand{\lam}{\lambda}
\newcommand{\dl}{\delta}

\newcommand{\Gm}{\Gamma}
\newcommand{\alV}{\nu}
\newcommand{\PR}{\mathrm{Pr}}

\newcommand{\me}[0]{\mathrm{e}}

\newcommand{\md}[0]{\mathrm{d}}

\begin{document}

\preprint{Bicocca-FT-03-25}
\title{The statistical geometry of scale-free random trees} 
\author{Luca Donetti} \email{Luca.Donetti@mib.infn.it} 
\author{Claudio Destri} \email{Claudio.Destri@mib.infn.it}
\affiliation{Dipartimento di Fisica
G. Occhialini, Universit\`a di Milano-Bicocca and INFN, Sezione di
Milano, Piazza delle Scienze 3 - I-20126 Milano, Italy}

\begin{abstract}
  The properties of random trees (Galton-Watson trees) with scale-free
  (power-like) probability distribution of coordinations are investigated
  in the thermodynamic limit. The scaling form of volume probability is
  found, the connectivity dimensions are determined and compared with other
  exponents which describe the growth. The (local) spectral dimension is
  also determined, through the study of the massless limit of the Gaussian
  model on such trees.
\end{abstract}

% 02.50.-r   Probability theory, stochastic processes, and statistics
% 05.40.-a   Fluctuation phenomena, rnadom processes, noise, and
%   Brownian motion
% 46.65.+g   Random phenomenon and media

\maketitle\tableofcontents

\section{Introduction and summary}

Graph theory and its applications play an important role
in many areas of scientific research, from pure mathematics, to physics,
statistics, biology and social sciences. In particular random graphs, that
is graphs extracted with some probability from a suitable statistical
ensemble, are interesting as a mean to implement the intrinsic complexity
and/or chaotic nature of many physical, biological and social systems \cite{nab}.

Among graphs, trees (that is graphs without loops) play a distinguished
role: they retain a deep interest and wide applicability while still being
amenable to detailed analytic studies. Random trees appears in several
distinct contexts, like polymer physics, critical percolation \cite{percol}
and two--dimensional quantum gravity (branched polymers \cite{branched}).

Generically, in a random graph the local coordination is itself a random
variable taking values according to some probability distribution; while in
the classical Erd\H{o}s-R\'enyi theory of random graphs \cite{bollobas}
this distribution is Poissonian, in several examples of ``experimental''
complex networks it turns out to be ``scale-free'', that is with a long
power-law tail \cite{barabasi}. This implies that on an infinite scale-free
graph nodes with diverging coordination are rather frequent, causing
some subtleties in the application of the law of large numbers.

In this paper we concentrate our attention to homogeneous scale-free
random trees subject only to the physically natural constraint of
being embeddable in a finite dimensional Euclidean space. This class
of random trees has been studied already in \cite{exotic,bck},
although from a viewpoint different from ours: while in these works
only some statistical averages relevant to characterize the geometry
of these trees were analyzed, we study here in depth the probability
distributions of the basic observables which specify the intrinsic
geometry. Our results apply therefore to any single ``generic''
(infinite) scale-free random tree and not just to their statistical
ensemble.

The outline of this paper is as follows. First of all we recall some
definitions of graph theory and we introduce the algorithm used to
build homogeneous random trees, summarizing known results. The local
connectivity dimension is then obtained from the scaling properties of
the probability distributions for the volume and the surface of the
spheres centered at the origin of the tree-constructing algorithm.
Details of the scaled probabilities are studied and their asymptotic
behavior is determined. Then the probability distributions for the
graph-averaged volume and the surface are considered; the average
connectivity dimension is extracted from their scaling properties.

Our results for the growth properties of a random tree can be
summarized in very concise heuristic formulas as follows: consider a sphere
(intrinsically defined in terms of the chemical distance alone) of radius
$r$ centered on some node; for large $r$ the volume $v$ of the sphere and
the coordination fluctuation $\Delta z$ within the sphere are random
variables simply related as
\begin{equation*}
  v \simeq (\Delta z)^2 r^2
\end{equation*}
This relation apply to all random trees with bounded growth rate (see
below) regardless of whether $\Delta z$ has a finite limit as $v\to\infty$
or not. In the former case, as it happens for instance on trees with
bounded coordination, one reads immediately out the connectivity dimension
$\dc=2$. In the latter case, one need to estimate the asymptotic dependence
of $\Delta z$ on $v$. For scale-free trees with tail exponent $2<\bt<3$
(see below for the proper definitions), one finds $\Delta z \sim
v^{(3-\bt)/(\bt-1)}$, yielding the $\bt-$dependent connectivity dimension
\begin{equation*}
  \dc = \frac{\bt-1}{\bt-2}
\end{equation*}
Suppose instead that one is interested in the graph-averaged volume
$\avg{v}$ obtained by averaging over all locations of the center of the
sphere (for a finite tree with large volume $V$). The random variable
$\avg{v}$ is related to the coordination fluctuation $\avg{\Delta} z$ {\em
  over the whole tree} just as before, that is
\begin{equation*}
  \avg{v} \simeq (\avg{\Delta} z)^2 r^2
\end{equation*}
Now the behavior in $r$ is always quadratic, regardless of the way the
coordinations are distributed. Of course, in the scale-free trees with
$2<\bt<3$ one has $\avg{\Delta} z \sim V^{(3-\bt)/(\bt-1)}$, making it
impossible to consider the standard thermodynamic limit $V\to\infty$.
Nonetheless, we find that a well defined limit exists for the
``renormalized'' volume $\avg{v}\,V^{(\bt-3)/(\bt-1)}$. Therefore, in all
cases we conclude for the average connectivity dimension
\begin{equation*}
  \dcb = 2
\end{equation*}
Finally, in the last section, the analysis of the probability distribution
for the effective squared mass, defined through the Gaussian model on the
tree, allows us to rigorously determine also the local spectral dimension
$\ds$. It fulfills the long-conjectured \cite{shlomo} relation
$\ds=2\dc/(\dc+1)$ already verified in \cite{bck} through a general
but completely different argument.

\section{Random trees}
\subsection{Generalities about graphs}

A graph $\G$ is defined by a set of nodes (finite or countable), pairwise
connected by a set of unordered links. If the set of nodes is finite, its
cardinality will be denoted by $|\G|$. The coordination number (or degree, or
simply coordination) of a node $x$ is the number of its nearest neighbors and
it is denoted by $z_x$. The intrinsic properties of the graph are determined
only by the connections between the nodes; the metric is given by the
so-called chemical distance: the distance $d(x,y)$ is equal to the number of
links of the shortest connected path between $x$ and $y$. This distance is
used to define spheres; the ``volume'' $v_r(x)$ is defined as the number of
nodes of the radius $r$ sphere centered at node $x$, and the ``surface''
$s_r(x)$ is the number of nodes of the corresponding spherical shell.  Here we
shall consider connected graphs in the limit of infinitely many nodes.

Given a function $F$ defined on the nodes, its average value $\avg{F}$ is
defined as the infinite radius limit of the average over spheres; if such
limit exists, it does not depend on the center of the spheres provided $F$
is bounded from below and the graph has a bounded growth rate \cite{rndtree},
that is the surface vanishes with respect to the volume in the infinite size
limit.

The properties of graphs can be described by various parameters; here
we consider the connectivity dimension \cite{suzuki} and the spectral
dimension \cite{alorb}. The connectivity dimension describes how the
volume of spheres scales with its radius $r$ for $r\to \infty$; we can
define a local connectivity dimension $\dc$ using the spheres centered
on any given node $x$
\begin{equation}\label{dc}
  v_r(x) \sim r^{\dc}
\end{equation}
It is not difficult to show that $\dc$ does not depend on $x$ when the
graph has a bounded growth rate. We can define also an average dimension
$\dcb$, if the average volume is used (provided the limit defining it
exists finite)
\begin{equation}\label{dcb}
  \avg{v_r} \sim r^{\dcb}
\end{equation}
These two connectivity dimensions usually coincide, but on strongly
inhomogeneous graphs they can be different \cite{combntd}.

The spectral dimension is related to long time properties of random walks
on the graph, as well as to many other physical properties such as the
infrared behavior of the Gaussian model \cite{hhw}. On a generic connected
graph $\G$ this model is defined by assigning a real-valued random variable
$\phi_x$ to each node $x\in \G$ with the Hamiltonian
\begin{equation}
  {\cal H} =   \frac12 \mu_0 \sum_x \phi_x^2 +
  \frac12 \sum_{\langle x,y \rangle} (\phi_x - \phi_y)^2  
\end{equation}
where $\langle x,y \rangle$ denotes nearest neighbors pairs of nodes and
$\mu_0>0$ is a free parameter (the squared mass in field--theoretic sense).
The spectral dimension is determined by the infrared $\mu_0\to0$ singularity
of the diagonal element of the covariance
\begin{equation}\label{ds}
     {\rm Sing}\, \gexpv{\phi_x^2} \sim  \mu_0^{\ds/2-1} 
\end{equation}
where $\gexpv{\cdot}$ denotes standard Gibbs expectation values weighted
with $\exp(-{\cal H})$. One can show that $\ds$ does not depend on the
choice of the node $x$ \cite{hhw}. As for the connectivity dimension, it is
possible to split the spectral dimension into a local one and an average
one. The definition in eq. \eqref{ds} evidently corresponds to the local
one; the average spectral dimension $\dsb$ characterizes in the same way
the singularity of the graph average of $\gexpv{\phi_x^2}$. Again, the two
dimensions usually coincide, but on strongly inhomogeneous graphs they can
be different \cite{combntd}.

Suppose now that a certain statistical ensemble $\Gens$ of (infinite)
graphs is given. As a consequence, connectivity and spectral dimensions
become in principle random variables with probability distributions
derived from the ``microscopic'' one specifying $\Gens$. Consider the
connectivity dimension: if the graphs of $\Gens$ are rooted, we can use the
root, denoted by $o$, as the center of the spheres those growth define
$\dc$; then the corresponding volumes $v_r(o)$ are random variables
distributed with some probability $\PR[v_r(o)=v]$. If this probability has a well
defined scaling behavior for $r\to\infty$, that is
\begin{equation*}
  \PR[v_r(o)=v] \simeq \frac1{r^d}\, p\!\left(\frac{v}{r^d}\right) 
\end{equation*}
then the random variable $v_r(o)r^{-d}$ has a well defined probability
distribution in the limit $r\to\infty$ and we may write again
\begin{equation*}
  v_r(o) \sim r^d
\end{equation*}
as in eq. \eqref{dc}, identifying $d$ with a non-fluctuating local
connectivity dimension $\dc$. In other words $\dc$ is a property of a
``generic'' specimen of the ensemble, that is a a property of $\Gens$
itself. A similar argument applies to the average connectivity dimension
$\dcb$ and to the spectral dimensions using the random variable
$\gexpv{\phi_x^2}$ and its graph average.
  
An alternative approach is used in ref. \cite{exotic}, where averages in
the canonical or grand-canonical ensembles of certain random trees ({\em
  i.e.} graphs without loops) are studied rather than a single ``generic''
sample tree. Two parameters are introduced to describe the intrinsic
geometry of these trees, also called ``branched polymers'': the Hausdorff
dimensions $d_H$ and $d_L$. The former is related to how the average
two-point distance on graphs with $V$ nodes scales with the size $V$:
\begin{equation} \label{dhdef}
  \expv{d(x,y)}_V \sim V^{1/d_H}  
\end{equation}
The latter (called local Hausdorff dimension in \cite{exotic}) is related
to the behavior of the two-points correlation function $g^{(2)}_V(r)$
(proportional to the number of couples of nodes which are at distance $r$)
for distances which are large, but much smaller than $V^{1/d_H}$:
\begin{equation} \label{dldef}
  g^{(2)}_V(r) \stackrel{V\to\infty}{\sim} r^{d_L-1} \; ,\qquad
   1\ll r\ll V^{1/d_H}
\end{equation}
It is claimed in \cite{exotic} that $d_H$ and $d_L$ differ on a certain
class of ``exotic'' random trees characterized by unbounded local
coordinations.

Actually, it is not a priori obvious if and how parameters such as
$d_H$ and $d_L$ are connected to the connectivity dimensions $\dc$ and
$\dcb$, although it is common lore to identify the Hausdorff
dimension with the connectivity dimension when no distinction is made about
local or average dimensions. One of the aims of the present work is just to
fully answer this question in the case of homogeneous random trees, as
we shall see below.

\subsection{Generalities about random trees}
Homogeneous random trees are built by the random independent
extraction of every node's degree from a given distribution $f_z$.
This process can be formulated as a Galton-Watson branching
process \cite{gwproc}; in particular, since we are interested in trees
with bounded growth rate, we must consider the critical Galton-Watson
case.  The given coordination distribution must be properly normalized
\begin{equation}\label{fnorm}
         \sum_z f_z = 1 
\end{equation}
and the average coordination must be equal to 2 because of the condition of
bounded growth rate (see ref. \cite{rndtree})
\begin{equation}\label{favg}
         \expv{z} = \sum_z z f_z = 2
\end{equation}
Clearly these sums have a finite number of addends if the coordination
is bounded, while they become series if it is unbounded as in the
scale-free case. The series, however, must be properly convergent in order
equations~(\ref{fnorm}) and~(\ref{favg}) to hold, so that $f_z$ must
vanish faster than $z^{-2}$ for $z\to \infty$.  Let us also introduce
the probability generating function $g(\lam)$
\begin{equation}\label{gdef}
 g(\lam)  = \sum_z  f_z \lam^{z-1} =
  f_1 + f_2 \lam +  f_3\lam^2 + \ldots
\end{equation}
which enjoys the properties
\begin{equation}\label{gprop}
         g(1)=1 \quad , \qquad g'(1)=1 
\end{equation}
easily derived from those of $f_z$.

A branching process effectively ends when a shell (that is a given
generation of sibling branches) is made only by nodes with coordination 1,
so that the next shell is empty. The trees produced in this way are all
finite because the surviving probability after $r$ generations vanishes for
$r\to\infty$ in the critical Galton-Watson process \cite{gwproc}. Since we
are interested in the thermodynamic limit we have two possibilities
\cite{rndtree}.  The first one is an explicit preconditioning on
non-extinction: this means that modified branching probabilities are used
to avoid finite trees, while keeping unaltered the properties of the
infinite trees \cite{kesten}.  This is achieved if the root coordination is
chosen with probability $f_z$ while on every successive shell the
coordination of the first node is extracted with a modified probability
distribution $\tf_z = (z-1) f_z$. The other possibility consists in
calculating probabilities conditioned on the number of nodes $V$ of the
resulting trees, and then take the limit $V\to\infty$. In this latter case
the root coordination must be extracted with the refined probability
\begin{equation}\label{fhat}
  \ftz = N_1 \frac{f_z}z \;,\quad  \frac1{N_1} = \sum_z \frac{f_z}z
\end{equation}
since the root has as many branches as its coordination while every other
node has one branch less than its coordination [see section 3.1 of
ref. \cite{rndtree} for details].

\subsection{Trees with bounded coordination}

In \cite{rndtree} and \cite{rndtree2} the geometrical and spectral
properties of bounded random trees were determined using non-extinction
preconditioning. The surface and volume probability, in the large radius
limit, are shown to be function only of the scaled variables $s/r$ and
$v/r^2$, respectively. Moreover, after the discussion of the auto-averaging
property, also average surface and volume are fully analyzed. These results
allow the rigorous determination of the connectivity dimension, both local
and average: $\dc = \dcb = 2$. The study of the Gaussian model on such
trees leads also to the determination of the spectral dimension $\ds=4/3$.
Using a simple scaling hypothesis it is shown that the relation
$\ds=2\dc / (\dc+1)$ should hold on a wider class of random trees.

An important universality property is also found: all the average
values and probability distributions in the large radius limit depend
on the $f_z$ distribution only through its second moment or,
equivalently, $g''(1)$. This parameter quantifies the
coordination fluctuations, since
\begin{equation*}
        g''(1) =  \sum_z f_z [z - \expv{z}]^2
\end{equation*} 
and characterizes all asymptotic behaviors.  Altogether, these results
can be summarized as
\begin{equation}\label{vtor}
  v \simeq g''(1)\,r^2
\end{equation}
meaning that $v/[g''(1)r^2]$, as $r\to\infty$, is a random variable with a
well defined universal probability distribution. 

The approach based of the grand-canonical ensemble of branched
polymers yields the result $d_H=d_L=2$ when the coordination of each
node is bounded. This corresponds to the ``generic phase'' of random
trees according to \cite{exotic}.

\subsection{Scale-free trees}

In this paper we turn to scale-free trees, that is trees whose
coordination distribution has a long power--law tail:
\begin{equation*}
  f_z \simeq A\, z^{-\bt} \ ,\qquad\text{for } z\to\infty
\end{equation*}
where $\bt>2$.  When the coordination is unbounded, the function
$g(\lam)$ may have singularities; in the scale-free case it becomes
singular for $\lam=1$. For any non-integer exponent\footnote{for
  integer $\bt$ the singularity has a different, logarithmic, form;
  however, since the final results will be analytic in $\bt$, one
  is able to extend them also to integer values of $\bt$.} its
expansion contains a term proportional to $(1-\lam)^{\bt-1}$, that is:
\begin{equation}
  \label{gexp}
  g(\lam) = g_a(\lam)+ \cs (1-\lam)^{\bt-1} 
\end{equation}
where $g_a(\lam)$ is analytic in $\lam=1$
\begin{equation}
  \label{gaexp}
  g_a(\lam) = 1 - (1-\lam) + c_2 (1-\lam)^2 +\ldots + c_k (1-\lam)^k +
  \ldots
\end{equation}
and $\cs = A\,\Gm(1-\bt)$. Therefore, if $2<\bt<3$, $g''(\lam)$ diverges for
$\lam\to1$; since we found $g''(1)$ to be an important parameter in the
bounded coordination case, we expect this divergence to have many important
consequences. As a first example it causes the divergence of $\expv{v_r(o)}$,
the expected volume of the balls around the root of the branching process.
This can be intuitively explained, using non-extinction preconditioning,
because on every shell the coordination of one node is chosen with probability
distribution $\tf_z$, whose first moment is infinite. If auto-averaging holds,
then also $\avg{v_r}$ would diverge on a single scale-free tree and the
definition \eqref{dcb} of average connectivity dimension appears troublesome
for scale-free trees. Thus, in order to deal with finite graph-averaged
quantities, we will also use the finite volume approach, regarding the volume
$V$ as a regulator.

We may then find an heuristic argument based on eq. (\ref{vtor}) to
determine the connectivity dimensions $\dc$ and $\dcb$ for scale-free
random trees. First of all, if $\bt>3$ nothing changes because
$g''(1)$ is still finite, so that $\dc=\dcb=2$. On the contrary, if
$2<\bt<3$, we have to face the divergence of $g''(1)$. But
coordination fluctuations are finite for a finite number $N$ of nodes,
and diverge with $N$ in a way fixed by the long tail of $f_z$, since
all coordination extractions are independent. Let $z_\mathrm{max}(N)$
be the largest coordination extracted, estimated from
\begin{equation*}
  \sum_{z=z_\mathrm{max}}^{\infty} \!\! f_z \sim z_\mathrm{max}^{1-\bt}
  = \frac1N 
\end{equation*}
then we may estimate fluctuations by
\begin{equation*}
  \sum_{z=1}^{z_\mathrm{max}(N)} \!\! (z-\expv{z})^2 f_z \sim
  [z_\mathrm{max}(N)]^{3-\bt} = N^{\alV}
\end{equation*}
where we defined the exponent $\alV$ as
\begin{equation*}
  \alV = (3-\bt)/(\bt-1)
\end{equation*}
Thus for the local growth we have the consistency relation
\begin{equation*}
  v_r \simeq v_r^{\alV} r^2
\end{equation*}
which implies
\begin{equation*}
  v_r \simeq  r^{2/(1-\alV)} = r^{(\bt-1)/(\bt-2)} \qquad \Rightarrow 
  \quad \dc = \frac{\bt-1}{\bt-2} 
\end{equation*}
When the graph-averaged $\avg{v_r}$ is considered for trees of $V$ nodes, the
coordination fluctuation over the whole graph should be used instead:
\begin{equation*}
  \avg{v_r} \simeq r^2 V^{(3-\bt)/(\bt-1)}
\end{equation*}
From this we read the scaling exponent $2$ independently on $V$, but
without the standard thermodynamic limit. Of course this is not a real
proof and a more rigorous approach confirming these results is adopted in
the following sections (see also Appendix \ref{sec:prob-distribution}); at
any rate, these exponents are the same as those calculated for scale-free
branched polymers with $2<\bt<3$ in the grand-canonical approach:
$d_H=(\bt-1)/(\bt-2)$ \cite{bck} and $d_L=2$ \cite{exotic}.

\section{Growth statistics}

\subsection{Scaling probabilities for local surface and volume}
\label{sv_probab}

The probability $\PR[s_r(o)=s]$ ($\PR[v_r(o)=v]$) for the surface (volume) of
the radius $r$ sphere around the root of an infinite tree can be found,
recursively on $r$, using the modified probabilities given by explicit
preconditioning (see \cite{rndtree}, where a slightly different notation is
used). The corresponding generating functions
\begin{gather*}
  G^s_r(\lam) = \sum_s \lam^s \, \PR[s_r(o)=s] \\
  G^v_r(\lam) = \sum_v \lam^v \, \PR[v_r(o)=v]
\end{gather*}
satisfy functional recursion rules easier to analyze, given by equations~(15)
and (16) in \cite{rndtree}, that is
\begin{equation*}
  \begin{split}
    G^s_{r+1}(\lam)  & = \lam\, g_{r+1}(\lam)\, g'_r(\lam) \\
    \frac{G^v_{r+1}(\lam,1)}{G^v_r(\lam,1)} & =
    \lam\,g'(h_{r-1}(\lam)) \,\frac{h_{r+1}(\lam)}{h_r(\lam)}
  \end{split}
\end{equation*}
where
\begin{align}
  \label{grrec}
  g_{r+1}(\lam) & = g(g_r(\lam)) & g_0(\lam) & = \lam \\
  \label{hrrec}
  h_{r+1}(\lam) & = \lam\,g(h_r(\lam)) & h_0(\lam) & = \lam 
\end{align}
These recursions involve only $g(\lam)$ and $g'(\lam)$, which are finite for
$0\le\lam\le1$ for every value of $\bt>2$, and therefore can be solved
numerically with high accuracy for any $r$ (subject only to computational
limits), thus providing a solution of our problem from a practical point
of view.
\begin{figure}[t]
  \centering 
  \psfrag{logv}[cc][cc]{\Large $\log(v)$} 
  \psfrag{logp}[cc][cc]{\Large $\log(p)$} 
  \psfrag{title}[cc][cc]{\Large $\bt=2.5,\quad r=10$} 
  \includegraphics[width=.8\textwidth]{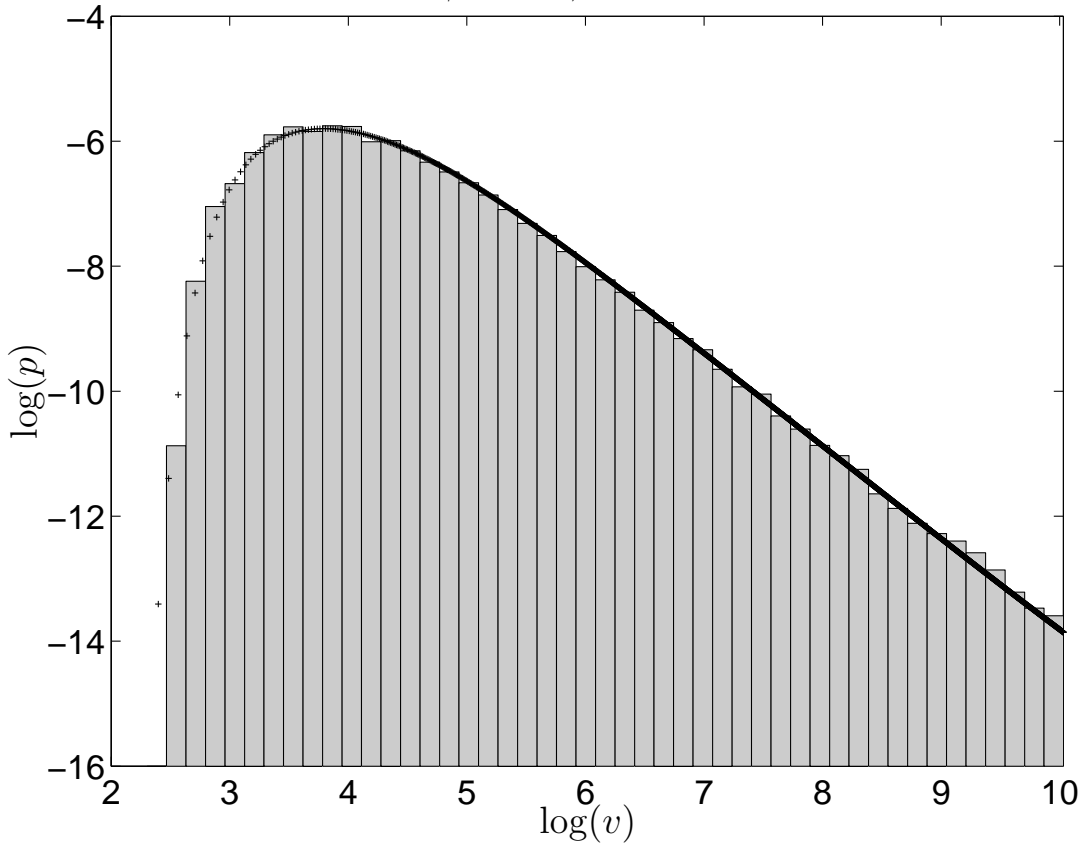}
  \caption{Numerically calculated volume distribution vs. simulation histogram}
  \label{f_distrib}
\end{figure}
In figure~\ref{f_distrib} the $r=10$ volume probability corresponding to a
tree with $\bt=2.5$ is plotted against the ``experimental'' frequency
distribution; this is obtained by sampling, over $1000$ different graph of
$8 \cdot 10^6$ nodes each, the volume of $2000$ balls centered on randomly
chosen nodes. The agreement is good, and we notice the power law tail with
exponent that can be estimated as $\simeq - 1.5$. In other cases (not
shown) this is always equal to $1-\bt$.

Analytically, upon substitution of eqs.~(\ref{gexp}) and~(\ref{gaexp}) for
$g(\lam)$, one can solve the recursions rules~(\ref{grrec}),~(\ref{hrrec})
for the first terms of the expansions of $g_r(\lam)$ and $h_r(\lam)$ as
$\lam\to1$:
\begin{equation}
  \label{grexp}
  \begin{split}
    g_r(\lam) & \simeq 1 - (1-\lam) + c_2 r (1-\lam)^2 + \ldots 
    + \cs r (1-\lam)^{\bt-1} \\
    h_r(\lam) & \simeq 1 + r (1-\lam) + \frac{c_2}3 r^3 (1-\lam)^2  +\ldots 
    + \frac{\cs}{\bt} r^{\bt} (1-\lam)^{\bt-1}
  \end{split}
\end{equation}
where only the leading order term in $r$ is shown for every power of
$(1-\lam)$. Similarly for $G^s_r(\lam)$ and $G^v_r(\lam)$ one finds
\begin{equation} 
  \label{Grexp}
  \begin{split}
    G^s_r(\lam) & \simeq 1 - 2 c_2 r (1-\lam) + \ldots 
    - (\bt-1) \cs r (1-\lam)^{\bt-2} \\
    G^v_r(\lam) &\simeq 1 - c_2 r^2 (1-\lam) +\ldots - 
    \cs r^{\bt-1} (1-\lam)^{\bt-2} 
  \end{split}
\end{equation}
Through the rules of (inverse) discrete Laplace transform, the
singular terms with noninteger powers in $1-\lam$ determinine power law
tails in the probability distributions at fixed $r$:
\begin{equation*}
  \PR[s_r(o)=s] \stackrel{s\to\infty}{\simeq} A\, r s^{-(\bt-1)}\;,\qquad 
  \PR[v_r(o)=v] \stackrel{v\to\infty}{\simeq} \frac{A}{\bt-1} 
  r^{\bt-1} v^{-(\bt-1)}
\end{equation*}
in agreement with numerical data.  The local connectivity dimension $\dc$ is
determined if we find scaling forms of these probabilities such that, as 
$s\to\infty$ or $v\to\infty$ {\em and} $r\to\infty$,
\begin{equation*}
  \PR[s_r(o)=s] \simeq \frac1{r^{\dc-1}}\,
  \phi\!\left(\frac{s}{r^{\dc-1}}\right) \;, \qquad 
  \PR[v_r(o)=v] \simeq \frac1{r^{\dc}}\,
  \Phi\!\left(\frac{v}{r^{\dc}}\right)
\end{equation*}
This means that $\dc$ must be such that $G^s_r(\exp(-u/r^{\dc-1})$ and
$G^v_r(\exp(-\xi/r^{\dc})$ have finite non-trivial limits for
$r\to\infty$. The right scaling exponent is found by examining
equation~(\ref{Grexp}): if $\bt>3$ the leading order term in $(1-\lam)$
is the linear one so that finite functions of $u$ and $\xi$ are
obtained with 
\begin{equation*}
  \dc=2 \qquad \qquad \text{for } \bt>3
\end{equation*}
while if $2<\bt<3$ the most important term is
the singular one so that the correct scaling is given by
\begin{equation}
  \label{dcl}
  \dc = \frac{\bt-1}{\bt-2} \qquad \qquad \text{for } 2<\bt<3
\end{equation}
In appendix~\ref{sec:prob-distribution} we present all explicit
calculations about the surface and volume probability distributions,
for the two cases $\bt>3$ and $\bt<3$, which confirm this simple
scaling analysis.

\subsection{Fixed volume expectation values}

In this section we study the local growth properties of scale-free random
trees of finite volume $V$. We shall reconstruct the dependence on $V$ of
the surface and volume probability distributions from the expectation
values of $s_r(o)$ and $v_r(o)$ and their powers. We are lead to consider
expectation values over finite-size trees, regarding their volume $V$
as a regulator, because the power-law tails with exponent $1-\bt$ found in
the previous section for $\PR[s_r(o)=s]$ and $\PR[v_r(o)=v]$
imply the divergence with $V$ of high enough moments of $s_r(o)$ or $v_r(o)$.

Let us observe that in this context the average is made over all possible
rooted trees with $V$ nodes; this can also be seen as the graph average
over all nodes of an unrooted tree, which is then further averaged over all
possible realization of the unrooted tree, that is
\begin{equation}\label{expvavg}
  \expv{O}_V = \expv{\avg{O}}_V
\end{equation}
for every observable $O$. In particular
\begin{equation*}
  \expv{s_r(o)^n}_V = \expv{\avg{s_r^n}\,}_V 
\end{equation*}
with a similar relation for the volumes. No dependence on the root $o$
survives and we shall drop it from the expectation values. 

Let us start by studying the probability $\Pt(V)$ that any rooted tree $T$
with $V$ nodes is produced by the branching process without the
non-extinction precondition
\begin{equation*}
  \Pt(V) = \PR[|T|=V]
\end{equation*}
By the tree-producing algorithm, $\Pt(V)$ satisfies the identity
\begin{equation*}
  \Pt(V) = \sum_z \ftz \!\! \sum_{V_1,\ldots,V_z} 
  \delta(V-1-\sum_{j=1}^z V_j) \prod_{j=1}^z \Pb(V_j)
\end{equation*}
where $\Pb(V)=\PR[|B|=V] $ is the same probability relative to a
branch $B$, that is a tree whose root has an incoming link (and $z-1$
branches) and whose coordination is extracted with probability $f_z$;
it satisfies a similar equation with $z-1$ instead of $z$ and $f_z$
instead of $\ftz$:
\begin{equation*}
  \Pb(V) = \sum_z f_z \!\! \sum_{V_1,\ldots,V_{z-1}} 
  \delta(V-1-\sum_{j=1}^{z-1} V_j) \prod_{j=1}^{z-1} \Pb(V_j)
\end{equation*}
The generating functions $\Gt(\lam)=\sum_V \Pt(V)\lam^V$ and
$\Gb(\lam)=\sum_V \Pb(V)\lam^V$ satisfy
\begin{equation}\label{Gb_eq}
  \Gb(\lam) = \lam \,g(\Gb(\lam)) \quad,\qquad \Gt(\lam) = \lam\, \ghat(\Gb(\lam))
\end{equation}
where $g(\lam)$ is the usual probability generating function defined
in equation~(\ref{gdef}), while
\begin{equation*}
  \ghat(\lam) = \sum_z \lam^z\, \ftz
\end{equation*}
Normalization of $\ftz$ implies $\ghat(1)=1$ and from the definition
easily follows that $\ghat(\lam)$ derivatives are proportional to
those of $g(\lam)$; more precisely we have
\begin{equation*}
  \ghat^{(k)}(\lam) = N_1 g^{(k-1)}(\lam)
\end{equation*}
Since we are interested in the $V\to\infty$ limit, which corresponds
to $\lam\to1$, we can use expansions~(\ref{gexp}) and (\ref{gaexp}) to
obtain the asymptotic behavior of $\Gb(\lam)$ and $\Gt(\lam)$. For
different values of $\bt$ we have different situations depending on
which term, the singular one or the quadratic one, is lower order. 

For $\bt>3$ we have
\begin{equation*}
  \begin{split}
    \Gb(\lam) & \simeq 1 - c_2^{-1/2} (1-\lam)^{1/2} \\
    \Gt(\lam) & \simeq 1 - N_1 c_2^{-1/2} (1-\lam)^{1/2}
  \end{split}
\end{equation*}
so that, in the $V\to\infty$, the probability $\Pt(V)$ reads
\begin{equation*}
  \Pt(V) \simeq N_1 \frac{1}{2\sqrt{c_2 \pi}} V^{-3/2}
\end{equation*}
For $2<\bt<3$, instead, the singular term produces
\begin{equation*}
  \begin{split}
    \Gb(\lam) & \simeq 1 - \cs^{-1/(\bt-1)} (1-\lam)^{1/(\bt-1)} \\
    \Gt(\lam) & \simeq 1 - N_1 \cs^{-1/(\bt-1)} (1-\lam)^{1/(\bt-1)} 
  \end{split}
\end{equation*}
Thus we have, for $V\to\infty$
\begin{equation}
  \Pt(V) \simeq N_1 \frac{-\cs^{-1/(\bt-1)}}{\Gm(-1/(\bt-1))} 
  V^{-\bt/(\bt-1)}  
\end{equation}

Let us now turn to the calculation of the expected size of a shell of
radius $r$ in trees with volume $V$. First of all let us introduce the
joint probability $\Pt_r(s,V)$ that the size of the tree is $V$ and
the $r$-th shell has $s$ nodes:
\begin{equation*}
  \Pt_r(s,V) = \PR[s_r(o)=s,|T|=V]
\end{equation*}
It is related to the same probability relative to branches $\Pb_r(s,V)$
as
\begin{equation*}
  \Pt_r(s,V) = \sum_z \ftz \sum_{V_1,\ldots,V_z} 
  \sum_{s_1,\ldots,s_z} 
  \delta(V-1-\sum_{j=1}^z V_j) \,\delta(s -\sum_{j=1}^z s_j) 
  \prod_{j=1}^z \Pb_{r-1}(s_j,V_j)
\end{equation*}
while $\Pb_r(s,V)$ satisfies a similar recursion which is obtained from the
previous one by replacing $z$ with $z-1$ and $\ftz$ with $f_z$,
\begin{equation}  \label{pbrec}
  \Pb_r(s,V) = \sum_z f_z \sum_{V_1,\ldots,V_{z-1}} 
  \sum_{s_1,\ldots,s_{z-1}} 
  \delta(V-1-\sum_{j=1}^{z-1} V_j) \,\delta(s-\sum_{j=1}^{z-1} s_j) 
  \prod_{j=1}^{z-1} \Pb_{r-1}(s_j,V_j)
\end{equation}
Clearly the expectation value of $s_r(o)$ on trees with $V$ nodes is now given
by
\begin{equation}
  \label{expv_s}
  \expv{s_r}_V = \frac{\sum_{s} s \Pt_r(s,V)}{\Pt(V)} =
  \frac{\Etvr{s}}{\Pt(V)}
\end{equation}
where $\Etvr{s}$ is defined as the weighted sum in the first
numerator. Analogously we can define $\Ebvr{s}$ as
\begin{equation*}
  \Ebvr{s} = \sum_{s} s\, \Pb_r(s,V)
\end{equation*}
and the two generating functions
\begin{equation*}
  \Etlr{s} = \sum_V \lam^V \Etvr{s} \; , \quad
  \Eblr{s} = \sum_V \lam^V \Ebvr{s} 
\end{equation*}
A recursion equation for the latter can be found using equation~(\ref{pbrec})
\begin{equation*}
  \begin{split}
    \Eblr{s} & = \sum_V \lam^V \sum_s  s\, \Pb_r(s,V) \\
    & = \lam \sum_z  f_z \sum_{V_1,\ldots,V_{z-1}} \sum_{s_1,\ldots,s_{z-1}}
    \Big(\sum_{k=1}^{z-1} s_k\Big) 
    \prod_{j=1}^{z-1} \lam^{V_j} \Pb_{r-1}(s_j,V_j) \\
    & = \lam \sum_z f_z (z-1) \Eblm{s} \Gb(\lam)^{z-2} \\
    & = \lam \, g'(\Gb(\lam)) \Eblm{s}
  \end{split}
\end{equation*}
and one for $\Etlr{s}$ is similarly obtained
\begin{equation*}
  \Etlr{s}  = \lam \, \ghat'(\Gb(\lam)) \Eblm{s} 
  = N_1 \lam \, g(\Gb(\lam)) \Eblm{s}    
\end{equation*}
Since $s_0(o)=1$, $\Eb{V}{0}{s}$ is equal to $\Pb(V)$, so that the
initial condition reads
\begin{equation*}
  \Eb{\lam}{0}{s} = \Gb(\lam)
\end{equation*}
Then, for $r\ge0$, the solutions read
\begin{equation}  \label{expv_sol}
  \begin{split}
    \Eblr{s} & = \Gb(\lam) \Big[\lam \, g'(\Gb(\lam)) \Big]^r \\
    \Etlr{s} & = N_1 \lam \, \Gb(\lam) g(\Gb(\lam))
    \Big[\lam \, g'(\Gb(\lam)) \Big]^{r-1}
  \end{split}
\end{equation}
Now $\expv{s_r}_V$ can be easily found, but we have to consider the
cases $\bt>3$ and $\bt<3$ separately.

For $\bt>3$ the $\lam\to1$ asymptotic expansion of $\Etlr{s}$ reads
\begin{equation*}
  \Etlr{s} \simeq N_1 \left[ 1 - (2+2 c_2(r-1)) c_2^{-1/2}
    (1-\lam)^{1/2} \right]
\end{equation*}
which imply the large $V$ limit of $\expv{s_r}_V$
\begin{equation*}
  \expv{s_r}_V \simeq 2+2 c_2(r-1)
\end{equation*}
For $2<\bt<3$ we have, instead,
\begin{equation*}
  \Etlr{s} \simeq N_1 \left[ 1 - (r-1) (\bt-1) \cs^{1/(\bt-1)}
  (1-\lam)^{(\bt-2)/(\bt-1)} \right]
\end{equation*}
and
\begin{equation} \label{expvs}
  \expv{s_r}_V \simeq k_1(\bt) (r-1) \,V^{\alV}
\end{equation}
where 
\begin{equation*}
  k_1(\bt)= (\bt-1)\cs^{2/(\bt-1)} 
  \frac{\Gm(-1/(\bt-1))}{\Gm(-(\bt-2)/(\bt-1))}  
\end{equation*}

In the first case the $V\to\infty$ limit is finite, while the second
one we find a divergence, as expected. To understand how
equation~(\ref{expvs}) relates with the known $\dc$, we now calculate
higher moments of $s_r$.  First of all we generalize the notation
$\Pt_r(O,V)$, $\Etvr{O}$ and $\Etlr{O}$ to every observable $O$
\begin{equation*}
  \begin{split}
    \Pt_r(O,V) & = \PR[O_r(x)=O,|T|=V] \\
    \Etvr{O} & = \sum_O O \Pt_r(O,V) \\
    \Etlr{O} & = \sum_V \lam^V \Etvr{O}
  \end{split}
\end{equation*}
together with their branch counterparts $\Pb_r(O,V)$, $\Ebvr{O}$ and
$\Eblr{O}$. Now it is easy to show that the recurrence rule for
$\Etlr{s^2}$ reads:
\begin{equation*}
  \Etlr{s^2} = \lam \, \ghat'(G(\lam)) \Eblm{s^2} +
  \lam\, \ghat''(G(\lam))\, (\Eblm{s})^2
\end{equation*}
Thus for $2<\bt<3$ the second moment is found to read
\begin{equation*}
  \expv{s^2_r}_V \simeq k_2(\bt) (r-1) \, V^{\alV + 1/(\bt-1)}
\end{equation*}
where $k_2(\bt)$ is some function of $\bt$ only.

For all higher moments $\Etlr{s^n}$ we find, as a general rule, that the
single $\Eblm{s^n}$ on the right hand side is multiplied by $\lam
\ghat'(G(\lam))$ while the products of $k$ $\Eblm{\cdot}$ are multiplied by
$\lam \ghat^{(k)}(G(\lam))$.  Moreover the equations for $\Eblr{O}$ are the
same as those for $\Etlr{O}$ with $\ghat(\cdot)$ substituted by $g(\cdot)$.
Therefore we obtain for all higher moments, up to the leading terms in the
$\lam\to1$ limit, 
\begin{equation}\label{srnrec}
  \begin{split}
    \Etlr{s^n} & \simeq \lam \, \ghat'(G(\lam)) \Eblm{s^n} 
    + \ldots \\
    \Eblr{s^n} & \simeq  \lam \, g'(G(\lam)) \Eblm{s^n} 
    + \lam \, g^{(n)}(G(\lam)) \Eblm{s}^n + \ldots    
  \end{split}
\end{equation}
This entails the final result
\begin{equation*}
  \expv{s^n_r}_V \simeq k_n(\bt) (r-1) V^{\alV + (n-1)/(\bt-1)}
\end{equation*}
which is compatible with a scaling form for the $s_r$ probability just given
by a finite-size scaling of the one found in section~\ref{sv_probab}, that is
\begin{equation*}
  \Pt_r(s|V) = \frac{\Pt_r(s,V)}{\Pt(V)} \simeq \frac1{r^{\dc-1}} \,
  q \! \left(\frac{s}{r^{\dc-1}}, \frac{r^{\dc}}{V} \right)
\end{equation*}
where $\dc$ is the already known local connectivity dimension
\begin{equation*}
  \dc = \frac{\bt-1}{\bt-2}
\end{equation*}
and the function $q(x,y)$ is such that
\begin{equation*}
  \int x^n q(x,y) \md x \simeq y^{(\bt-2-n)/(\bt-1)}
\end{equation*}

Similarly it is possible to calculate the finite volume moments of the
volume $v_r(o)$ which read:
\begin{equation}
  \label{expvn}
  \expv{v_r^n}_V \sim r^{n+1} \,V^{(n+2-\bt)/(\bt-1)}
\end{equation}
\begin{figure}[t]
  \centering 
  \psfrag{v/rdc}[cc][cc]{$v/r^{\dc}$} 
  \psfrag{rdc/V}[cc][cc]{$r^{\dc}/V $} 
  \psfrag{title}[cc][cc]{$\bt=8/3,\quad \dc=5/2,\quad \text{integrated probability:}\ 10\%\ 20\%\ 40\%\ 60\%\ 80\%\ 90\%$} 
  \includegraphics[width=.8\textwidth]{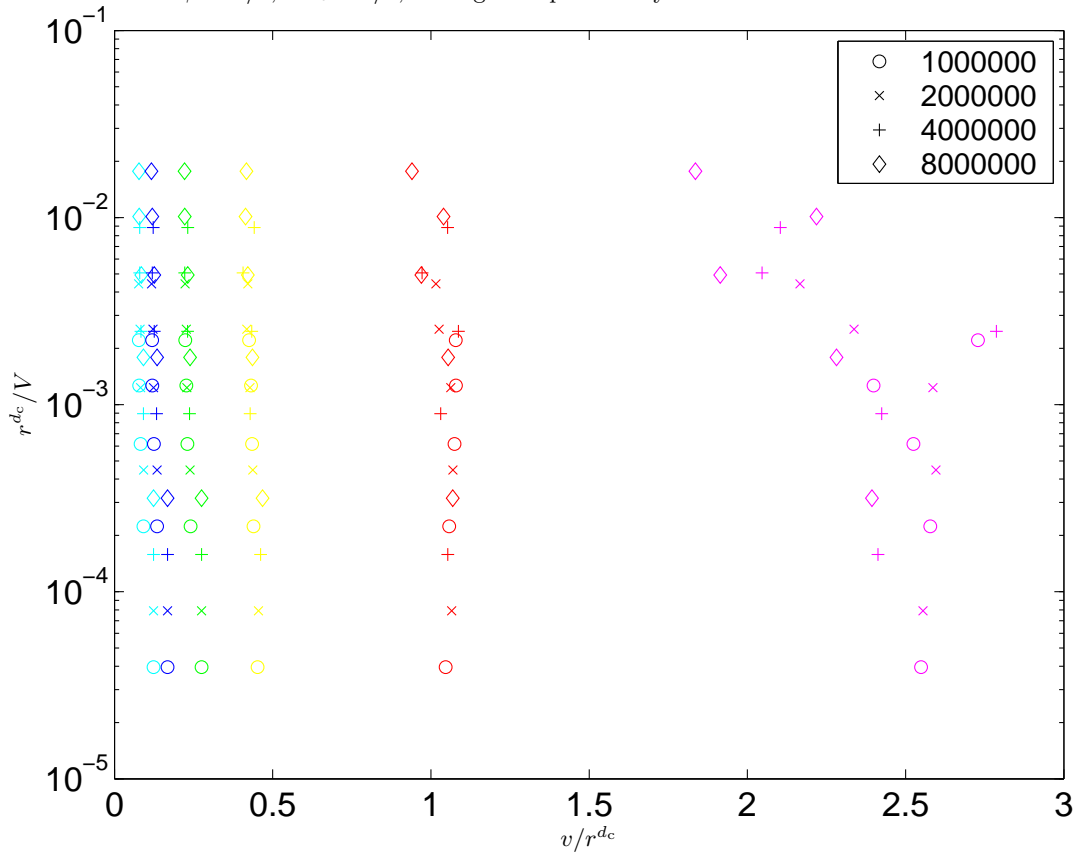}
  \caption{Group of symbols of the same color (online version) or shade of
    gray (printed version) correspond to all possible combinations of
    $r=10,20,30,40,50$ (using the same symbol) and $V=1,2,4,8\cdot 10^6$
    (using different symbols as explained in the legend).}
  \label{f_scaling}
\end{figure}
The proper scaling compatible with this result is
\begin{equation}\label{pscaling}
  \Pt_r(v|V) \simeq \frac1{r^{\dc}} 
  p\! \left(\frac{v}{r^{\dc}}, \frac{r^{\dc}}{V} \right)
\end{equation}
if the moments of the scaling function $p(x,y)$ behave as
\begin{equation*}
  \int x^n p(x,y) \md x \simeq y^{(\bt-2-n)/(\bt-1)}
\end{equation*}

Scaling~(\ref{pscaling}) agrees with numerical simulations as shown in
figure~\ref{f_scaling}. For several different values of $r$ and $V$, we
built 720 trees and calculated the volume $v$ of radius $r$ balls around
1000 (for each tree) randomly chosen nodes. Thus different ``experimental''
probability distributions are found; we considered the integrated
probability $Q_r(v|V)=\sum_{v'\le v}\Pt_r(v|V)$ and plotted the values of
$v$ corresponding to $Q_r(v|V)=0.1,\,0.2,\,0.4,\,0.6,\,0.8,\,0.9$; (after
the proper rescaling by $r^{\dc}$) versus $r^{\dc}/V$. For everyone of the
six values of $Q_r(v|V)$, the points corresponding to different values of
$r$ and $V$ fall roughly on a curve, apart from statistical errors
(fluctuations are expecially large for $Q_r(v|V)=0.9$, due to the long
power-law tail of $\Pt_r(v|V)$), so that
the form~(\ref{pscaling}) of the probability scaling is supported.

\subsection{Probability of the average surface}
\label{avg-conn-dim}

In the bounded coordination case we were able to prove the auto-averaging
property for every local observable: the graph average $\avg{O}$ is
non-fluctuating on infinite trees and coincides with the expectation value
$\expv{O}$ in the branching process around the origin
\cite{rndtree,rndtree2}. This implies immediately that local and average
connectivity dimensions coincide. 

In the scale-free case the situation is more involved because many relevant
expectation values do not even exist on infinite trees, so that we cannot
prove the auto-averaging property in the same way. Thus, to determine the
average connectivity dimension $\dcb$ we will try to directly investigate
the probability distribution of the average surface $\avg{s_r}$ in the
large $r$ limit. In particular, we look for a scaling form of the type
\begin{equation*}
  \PR[\avg{s_r}= s] \simeq \frac1{r^{\dcb-1}}
  f \! \left(\frac{s}{r^{\dcb-1}}\right)
\end{equation*}
A very close approach is possible also for the average volume $\avg{v_r}$,
but it is more involved and will not be reported here.

The results of the previous section, together with
equation~(\ref{expvavg}), allow us to calculate only the first moment
of such probability distribution. We now reconstruct it (at least its
scaling form) through the calculation of higher moments.

This can be done by introducing $S_r$, defined as the sum of $s_r(x)$
over all nodes $x$ of a tree:
\begin{equation*}
  S_r = \sum_{x\in T} s_r(x)
\end{equation*}
Then on a finite-sized tree, the average surface $\avg{s_r}$, is
given by
\begin{equation*}
  \avg{s_r} = \frac{S_r}{V}
\end{equation*}
and its (finite volume) moments are simply obtained
\begin{equation*}
  \expv{(\avg{s_r})^n}_V = \frac{\expv{S_r^n}_V}{V^n} 
\end{equation*}

The recursive nature of the trees allows us again to find the composition
rule for $S_r$ in terms of the values of $S_{r,j}$ and the local surface
$s_{k,j}$ (with $k<r$) relative to $j-$th branch around the root $o$. It
reads
\begin{equation} \label{Srcomp}
  S_r = \sum_j \left[S_{r,j} + 2 s_{r-1,j} + \sum_{k=0}^{r-2}
  \sum_{j'\neq j}  s_{k,j} s_{r-k-2,j'} \right]
\end{equation}
The recursion rule for the joint probability of
$S_r,s_1,\ldots,s_{r-1}$ and $V$ easily follows from (\ref{Srcomp})
and the recursion rules for $s_r$ and $V$.
With calculations similar to those of the previous section one obtains
\begin{equation*}
  \Etlr{S} = \lam\, \ghat'(G(\lam)) \Eblr{S} 
  + 2 \lam\, \ghat'(G(\lam))\Eblm{s} 
  + \lam\, \ghat''(G(\lam)) \sum_{k=0}^{r-2} \Eb{\lam}{k}{s} 
  \Eb{\lam}{r-k-2}{s}
\end{equation*}
and, as usual, the equation for $\Eblr{S}$ which is equal to this one
with all $\ghat(\cdot)$ substituted by $g(\cdot)$.  The expectation
value
\begin{equation*}
  \expv{\avg{s_r}}_V = \frac{\Etlr{S}}{V \Pt(V)}
\end{equation*}
is then obtained by straightforward calculations in the $V\to\infty$
limit, and the result is equal to the large $V$ limit of
$\expv{s_r}_V$, as expected [recall eq. (\ref{expvavg})].

The differences of course emerge when higher moments are considered, since
$\avg{s_r^n}\neq (\avg{s_r})^n$ for $n>1$.  $\expv{(\avg{s_r})^2}_V$ can be
obtained using the equations (only the leading terms are displayed)
\begin{equation*}
  \begin{split}
    \Etlr{S^2} \simeq & \lam \, \ghat'(G(\lam)) \Eblr{S^2} + \ldots \\
    \Eblr{S^2} \simeq & \lam \, g'(G(\lam)) \Eblr{S^2} + 
    \lam \, g''(G(\lam)) \Eblr{S}^2 + \\ 
    &+ \lam \, g^{(4)}(G(\lam)) \sum_{k=0}^{r-2} \sum_{k'=0}^{r-2} 
    \Eb{\lam}{k}{s} \Eb{\lam}{k'}{s} 
    \Eb{\lam}{r-k-2}{s} \Eb{\lam}{r-k'-2}{s} + \ldots
  \end{split}
\end{equation*}
and turns out to read
\begin{equation*}
  \expv{(\avg{s_r})^2}_V \simeq 
  \begin{cases}
    (2+2 c_2(r-1))^2 & \text{for } \bt>3 \\
    {\bar k}_2(\bt) (r-1)^2 V^{2 \alV} & \text{for } 2<\bt<3 \\
  \end{cases}
\end{equation*}
where ${\bar k}_2(\bt)$ does not depend on $r$ or $V$.

Then all the following moments can be calculated through the equations
\begin{equation}\label{Srnrec}
  \begin{split}    
    \Etlr{S^n} &\simeq \lam \, \ghat'(G(\lam)) \Eblr{S^n} + \ldots \\
    \Eblr{S^n} &\simeq \lam \, g'(G(\lam)) \Eblr{S^n}
    + \lam \, g''(G(\lam)) \frac12 \sum_j \frac{n!}{j!(n-j)!}
    \Eblr{S^j} \Eblr{S_r^{n-j}} + \\
    & + \lam \, g'''(G(\lam)) \frac1{3!}\sum_{j,j'}
    \frac{n!}{j!j'!(n-j-j')!} \Eblr{S^j} \Eblr{S^{j'}}
    \Eblr{S^{n-j-j'}} + \\
    & + \ldots + \lam \, g^{(n)}(G(\lam)) \Eblr{S}^n + \\
    & + \lam \, g^{(2n)}(G(\lam)) \sum_{k_1,\ldots,k_n} \Eb{\lam}{k_1}{s}
    \Eb{\lam}{r-k_1-2}{s} \cdots \Eb{\lam}{k_n}{s}
    \Eb{\lam}{r-k_n-2}{s} + \ldots
  \end{split}
\end{equation}
From the leading singularity in $1-\lam$ we read their asymptotic behavior
for large $V$:
\begin{equation}\label{sbarn}
  \expv{(\avg{s_r})^n}_V \simeq 
  \begin{cases}
    (2+2 c_2(r-1))^n & \text{for } \bt>3 \\
    {\bar k}_n(\bt) (r-1)^n V^{n \alV} & \text{for } 2<\bt<3 \\
  \end{cases}
\end{equation}

For $\bt>3$, since all moments are powers of the same finite quantity,
the probability distribution for $\avg{s_r}$ must be a delta function
in the $V\to\infty$ limit
\begin{equation*}
  \PR[\avg{s_r}= s] = \dl(s - (2+2 c_2(r-1))
  = \dl(s - \expv{s_r})
\end{equation*}
Thus in the thermodynamic limit the average does not fluctuate and
coincides with the expectation value for the surface around the root
of the tree and the auto-averaging property is verified for the observable
$s_r$. In particular we can extract the average connectivity dimension 
\begin{equation*}
  \dcb = \dc = 2
\end{equation*}
Similarly the auto-averaging property can be verified for the volume;
we expect it to hold in general for every observable with a finite
expectation value.

When $2<\bt<3$ instead, from eq.~(\ref{sbarn}) we can read the following
scaling form for the average surface probability
\begin{equation} \label{scalavgs}
  \Pt_r(\avg{s}|V) \simeq \frac1{r V^{\alV}}
  \overline{q}\!\left(\frac{\avg{s}}{r V^{\alV}} \right)
\end{equation}
This shows that for every $V$ there exists a scaling form of the probability
distribution of $\avg{s}$ for all scale-free trees of size $V$, but it does
not have a non-trivial standard thermodynamic limit. Although $\avg{s_r}$
diverges on infinite trees and the standard definition of
average connectivity dimension cannot be applied, we see that a
$V$-independent probability exists for the ``renormalized''
average surface $\avg{s}V^{-\alV}$ and that this scales according to the
average connectivity dimension $\dcb=2$.

For finite trees a comparison can be made with the definition of $d_L$
(equation~(\ref{dldef})): since the average surface differs from the two
point function only through a normalization factor
\begin{equation*}
   g^{(2)}_V(r) \propto \expv{s_r}_V \propto r
\end{equation*}
our calculation not only confirms the value $d_L=2$; it also shows
that this is not a property of the first moment only, but it holds  
``in probability'', that is for every ``generic'' scale-free random tree.

\section{Spectral dimension}
\label{sec:spectral-dimension}

In order to find the (local) spectral dimension of scale-free trees, we may
repeat the approach of \cite{rndtree2} substituting fixed radius averages
with fixed volume averages.  The quantity $\gexpv{\phi^2_x}$ is given by a
normalized Gaussian integral over all variables $\phi_y, y\in G$; if we
perform all integrations except the one over $\phi_x$ we are left with a
last integral which is also Gaussian and normalized, thanks to the
self-reproducing property of Gaussian integrals. Therefore we can define
the effective squared mass $\mu(x)$ from the width of this last integral:
\begin{equation*}
  \gexpv{\phi^2_x} = \sqrt{\frac{\mu(x)}{2\pi}}
  \int \md\phi_x\, \phi_x^2 \,\me^{-\mu(x) \phi_x^2/2} = \frac1{2\mu(x)}
\end{equation*}
On a tree produced by a branching process the rules of Gaussian
integration allow us to express the effective squared mass of the root
as a function of those of the branches \cite{rndtree2}.
\begin{equation}
  \label{murec}
  \mu(x) = \mu_0 + \sch{x}{y} \frac{\mu(y)}{1+\mu(y)}
\end{equation}
Similarly the effective squared mass of a branch can be expressed as a
function of those if its sub-branches, in a recursive way.
%% We then look for a scaling form for the average effective squared
%% mass:
%% \begin{equation}
%%   \label{muscaling}
%%   \expv{\mu}_V \simeq \mu_0 V^\dl F(\mu_0 V^\gm) 
%% \end{equation}
Since the recursion rule for $\mu(x)$ is highly nontrivial we are not
able to find directly its probability distribution or its moments;
instead we expand $\mu(x)$ in powers of $\mu_0$ and we consider the
coefficients $\V{n}$
\begin{equation*}
  \mu(x) = \sum_{n=1}^\infty (-1)^{n+1} \V{n}(x) \mu_0^n
\end{equation*}
Substituting this expansion in equation~(\ref{murec}), recursion rules
for the cofficients are obtained:
\begin{gather*}
  \V1(x) = 1 + \sch{x}{y} \V1(y) \\
  \V2(x) = \sch{x}{y} \Big[\V2(y) + \V1(y)^2\Big] \\
  \intertext{and, in general}
  \V{n}(x) = \delta_{n,1} + \sch{x}{y}\big[\V{n}(y) + 
  \cF_n(\V1(y),\V2(y),\ldots,\V{n-1}(y)) \big]
\end{gather*}
where $\cF_n$ is explicitly given in appendix~\ref{sec:vn-averages}.
The important feature which allows the calculation of $\V{n}$ moments
is that the expression of $\V{n}(x)$ is linear in the $\V{n}(y)$. 

Clearly $\V{1}$ is just the volume $V$ so that
\begin{equation*}
  \expv{\V1}_{V} = V \qquad \text{and} \qquad \expv{\V1^n}_{V} = V^n
\end{equation*}
Now we need the averages $\expv{\V{n}}_V$ for $n\ge 2$.
We use the same notation as in the previous sections ($\Pt(\cdot)$,
$\Etv{\cdot}$, $\Etl{\cdot}$, $\Pb(\cdot)$, \ldots ) with the only
difference that now observables do not depend on $r$.

% The recursion rule for the joint probability of $\V1$ and $\V2$:
% \begin{equation*}
%   \Pb(\V1,\V2) = \! \sum_z f_z \!\!\!\!  \sum_{\VV11,\ldots,\VV1{z-1}} 
%   \sum_{\VV21,\ldots,\VV2{z-1}} \!\!\!\! \dl\Big(\V1-1-\sum_j \VV1j \Big)
%   \dl\Big(\V2-\sum_j \big(\VV2j+\VV1j^2\big)\Big) \prod_j \Pb(\VV1j,\VV2j)
% \end{equation*}
% The equations for $\Ebv[\V1]{\V2}$ and its generating function
% $\Ebl{\V2}$ easily follow
% \begin{equation}
%   \label{avgvv}
%   \begin{split}
%     \Ebv[\V1]{\V2} & = \sum_{\V2} \V2 \Pb(\V1,\V2) \\
%     & = \sum_z f_z \!\! \sum_{\VV1j,\VV2j} \!\! 
%     \dl\Big(\V1-1-\sum_j \VV1j\Big)
%     \sum_j \Big[\VV2j+\VV1j^2\Big] \prod_j \Pb(\VV1j,\VV2j) \\
%     & = \sum_z f_z \sum_{\VV1j} \dl\Big(\V1-1-\sum_j \VV1j\Big) 
%     \sum_j \Big[\Ebv[\VV1j]{V_2}+\VV1j^2 \Pb(\VV1j)\Big]
%     \prod_{j'\neq j} \Pb(\VV1{j'}) \\
%     \Ebl{\V2} & = \lam \sum_z f_z (z-1) \Big[ \Ebl{\V2} +
%     \Ebl{\V1^2} \Big] \Gb(\lam)^{z-2} \\
%     & = \lam g'(\Gb(\lam)) \Big[ \Ebl{\V2} + \Ebl{\V1^2}  \Big] \\
%     & = \Big[1-\lam g'(\Gb(\lam)\Big]^{-1} \lam g'(\Gb(\lam))
%     \Ebl{\V1^2}    
%   \end{split}
% \end{equation}
Let us start with the expectation value of $\V2$; equations for
$\Etl{\V2}$ and $\Ebl{\V2}$ can be easily obtained in the same way as
before
\begin{equation}
  \label{avgvv}
  \begin{split}
    \Etl{\V2} & 
    = \lam \, \ghat'(\Gb(\lam)) \Big[ \Ebl{\V2} + \Ebl{\V1^2} \Big] \\
    \Ebl{\V2} & = \lam \, g'(\Gb(\lam)) \Big[ \Ebl{\V2} + \Ebl{\V1^2}  \Big] \\
    & = \Big[1-\lam \, g'(\Gb(\lam))\Big]^{-1} \lam \, g'(\Gb(\lam))
    \Ebl{\V1^2}
  \end{split}
\end{equation}
This is valid for any $f_z$ distribution, but now the $2<\bt<3$ and
$\bt>3$ cases must be examined separately.

When $2<\bt<3$ we can read the asymptotic behavior of $\expv{\V2}$ from the
leading $1-\lam$ singularity in equation~(\ref{avgvv}):
\begin{equation*}
  \expv{\V2}_V = \frac{\Etv{\V2}}{\Pt(V)} \sim
  V^{(3\bt-4)/(\bt-1)} \quad,\qquad V\to\infty
\end{equation*}
In a similar way it can be proved that (see
appendix~\ref{sec:vn-averages} for details):
\begin{equation}
  \label{expVn}
  \expv{\V{n}}_V = \frac{\Etv{\V{n}}}{\Pt(V)} 
  \sim V^{1+(n-1)(2\bt-3)/(\bt-1)} \quad,\qquad V\to\infty
\end{equation}
We can then write
\begin{equation}
  \label{mu_scaling}
  \expv{\mu}_V \simeq \mu_0 V F_1\Big(\mu_0 V^{(2\bt-3)/(\bt-1)}\Big) 
\end{equation}
The existence of the thermodynamic limit\cite{hhw} requires that
$F_1(t) \sim t^{-(\bt-1)/(2\bt-3)}$ for $t\to\infty$ so that the
powers of $V$ cancel out and a finite limit is obtained. After the
$V\to\infty$ limit, for $\mu_0\to0$ we have (see section 5 of
\cite{rndtree2} for a discussion on the order of the limits)
\begin{equation*}
  \expv{\mu}_\infty = \lim_{V\to\infty}\expv{\mu}_V 
  \propto \mu_0^{(\bt-2)/(2\bt-3)} \quad,\qquad \mu_0\to0
\end{equation*}
Equation~(\ref{expvVn}) in appendix~\ref{sec:vn-averages} can be used
to show that the higher moments of $\mu$ follow similar scaling laws:
\begin{equation*}
  \expv{\mu^n}_V \simeq \mu_0^n V^n F_n\Big(\mu_0 V^{(2\bt-3)/(\bt-1)}\Big) 
\end{equation*}
so that 
\begin{equation*}
  \expv{\mu^n}_\infty \sim \mu_0^{n (\bt-2)/(2\bt-3)} 
  \quad,\qquad \mu_0\to0
\end{equation*}
Since all moments are proportional to powers of the same quantity, we
can say that there exists a limit probability distribution for the
scaled variable $\mu \mu_0^{-(\bt-2)/(2\bt-3)}$ so that the local
spectral dimension reads:
\begin{equation*}
  \ds = 2 \frac{\bt-1}{2\bt-3} \quad \qquad \text{for } 2<\bt<3
\end{equation*}

When $\bt>3$ equation ~(\ref{avgvv}) allows us to write
\begin{equation*}
  \expv{\V2}_V = \frac{\Etv{\V2}}{\Pt(V)} \sim
  V^{5/2}  \quad,\qquad V\to\infty
\end{equation*}
All other moments are calculated in appendix~\ref{sec:vn-averages}
and, following the same steps as before, we can write
\begin{equation*}
  \expv{\mu^n}_\infty \sim \mu_0^{n/3} \quad,\qquad \mu_0\to0
\end{equation*}
so that 
\begin{equation*}
  \ds = \frac43 \quad \qquad \text{for } \bt>3
\end{equation*}
as in the bounded coordination case\cite{rndtree2}.  

Finally we notice that in both cases the relation
\begin{equation*}
  \ds = 2 \frac{\dc}{\dc+1} 
\end{equation*}
between the connectivity and the spectral dimension is fulfilled.

\appendix

\section{Details on probability distributions}
\label{sec:prob-distribution}
In order to find the probability distributions for surface and volume
we can follow the same steps of \cite{rndtree}. If the proper scaling
form of the surface and volume generating functions are substituted in
their recursion rules, differential equations are obtained by
consistency requirements. The solutions are the Laplace transform of
the scaled surface and volume probability distributions. 
For $\bt>3$ both Laplace transforms can be calculated, while for
$2<\bt<3$ only the surface probability Laplace transform and the
asymptotic behavior of the volume one can be obtained.

\subsection{$\bt>3$}

When $\bt>3$, $\dc$ is equal to $2$ and we can write $g_r(\lam)$ as
\begin{equation}
  \label{grexpA}
  g_r\big(\me^{-u/r}\big) 
  = 1 - \frac{a(u)}{r} + \cdots + \frac{a_*(u)}{r^{\bt-2}} + \cdots
\end{equation}
with the condition
\begin{equation}
  \label{a_condA}
    a(0) = 0\;, \qquad a'(0)  = 1
\end{equation}
that follows from $g_r(1)=g_r'(1)=1$ and
\begin{equation}
  \label{as_condA}
  a_*(u) \simeq \cs u^{\bt-1} \qquad \text{for } u\to 0
\end{equation}
derived from equation~(\ref{grexp}). This expansion imply those of
$G^s_r(\lam)$ and $\PR[s_r(o)=s]$:
\begin{gather}
  G^s_r\big(\me^{-u/r}\big) = a'(u) + \ldots - \frac{a_*'(u)}{r^{\bt-3}} +
  \ldots \nonumber \\
  \label{PrexpA}
  \PR[s_r(o)=s] = \frac1r \, \phi\!\left(\frac{s}r\right) + \ldots +
  \frac1{r^{\bt-2}} \, \phi_*\!\!\left(\frac{s}r\right) + \ldots
\end{gather}
where $\phi(x)$ and $\phi_*(x)$ are the inverse Laplace transforms of
$a'(u)$ and $a_*'(u)$, respectively. The analytic term $\phi(x)$ is the
same as in the bounded case; the singular term $\phi_*(x)$, even if it is
suppressed by powers of $r$, is essential in order to reproduce the
power-law tail.  The functions $a(u)$ and $a_*(u)$ are determined by
inserting equation~(\ref{grexpA}) in the recurrence~(\ref{grrec}), and
equating the coefficients of the corresponding powers of $r$. Thus two
coupled differential equations are obtained (see ref. \cite{rndtree} for
details on the method in the case of bounded trees, when only $a(u)$ is
required):
\begin{equation*}
  \begin{split}
    u a'(u) & = a(u) - c_2 a(u)^2 \\
    u a_*'(u) & = (\bt-2) a_*(u) - 2 c_2 a(u) a_*(u) +
    \cs a(u)^{\bt-1}
  \end{split}
\end{equation*}
The solution of the former, obtained with the use of the initial
condition (\ref{a_condA}), reads
\begin{equation*}
  a(u) = \frac{u}{1+c_2 u}
\end{equation*}
while the general solution of the latter reads
\begin{equation*}
  a_*(u) = \frac{\cs}{c_2 (\bt-4)} u^{\bt-2} 
  \left[ \frac{k}{(1+c_2 u)^2} - \frac1{(1+c_2 u)^{\bt-2}} \right]
\end{equation*}
The value of $k$ is found by comparison with the
condition~(\ref{as_condA}) and turns out to be $k=1$.  The inverse
Laplace transforms $\phi(x)$ of $a'(u)$ is
\begin{equation*}
  \phi(x) = \frac{x}{c_2^2} \me^{-x/c_2}
\end{equation*}
while only the asymptotic $x\to\infty$ behavior of $\phi_*(x)$ can be
explicitly calculated:
\begin{equation*}
  \phi_*(x) \simeq A\, x^{-(\bt-1)} 
\end{equation*}
Therefore, even if the contribution of $\phi_*(x)$ in
equation~(\ref{PrexpA}) is suppressed by powers of $r$, it is the most
important one for large $x$ because of its power-law tail compared
with the exponential one of $\phi(x)$. Moreover, if $\bt>4$ there are
other lower order corrections before the singular one; however since their
contribution in the Laplace transform $G^s_r(\exp(u/r))$ is analytic
for $u\to0$, they are exponentially vanishing for large $s/r$.
The large $s$ tail is therefore due to the singular term. 

Similar calculations can also be done for the volume probability. The
first step consists in writing the expansion of $h_r(\lam)$ with the
proper scaling:
\begin{equation}
  \label{hrexpA}
  h_r\big(\me^{-\xi/r^2}\big) = 1 - \frac{b(\xi)}{r} + \cdots +
  \frac{b_*(\xi)}{r^{\bt-2}} + \cdots
\end{equation}
with the condition
\begin{equation}
  \label{b_condA}
  b(0) = 0\;, \qquad b'(0) = 1
\end{equation}
that comes from $h_r(1)=1$ and $h_r'(1)=r$ and
\begin{equation}
  \label{bs_condA}
  b_*(\xi) \simeq \frac{\cs}{\bt} u^{\bt-1}
\end{equation}
derived from equation~(\ref{grexp}).  Inserting
equation~(\ref{hrexpA}) in the recurrence for $h_r(\lam)$
(equation~(\ref{hrrec})) and equating the coefficients of
corresponding powers of $r$, we obtain:
\begin{equation*}
  \begin{split}
    2 \xi b'(\xi) & = b(\xi) - c_2 b(\xi)^2 + \xi\\
    2 \xi b_*'(\xi) & = (\bt-2) b_*(\xi) - 2 c_2 b(\xi) b_*(\xi) 
    + \cs b(\xi)^{\bt-1}
  \end{split}
\end{equation*}
Their solutions, imposing the conditions~(\ref{b_condA})
and~(\ref{bs_condA}), can be found to read:
\begin{equation*}
  \begin{split}
      b(\xi) & = \sqrt{\frac{\xi}{c_2}} \tanh \sqrt{c_2 \xi}  \\
      b_*(\xi) & = \frac{\cs}{2 \xi} 
      \left(\sqrt{\frac{\xi}{c_2}} \tanh \sqrt{c_2 \xi} \right)^{\bt} 
      \left[ 1 - \frac{\bt-2}{\bt}
        \frac{{}_2F_1\!\left(1, \bt/2;1+\bt/2; 
            (\tanh \sqrt{c_2 \xi})^2\right)}{ (\cosh \sqrt{c_2 \xi})^2}
      \right] 
  \end{split}
\end{equation*}
where ${}_2F_1(a,b;c;z)$ is the hyper-geometric function; in our case 
it can also be written as:
\begin{equation} \label{hyperg}
  {}_2F_1\!\left(1,b;1+b; z \right) =
  \sum_{n=0}^{\infty} \frac{b}{b+n} z^n
\end{equation}

The volume probability $\PR[v_r(o)=v]$ and its generating function
$G^v_r(\lam)$ now can be written as 
\begin{gather*}
  G^v_r\big(\me^{-\xi/r^2}\big) = l(\xi) \left( 1 + \ldots -
    \frac{l_*(\xi)}{r^{\bt-3}} + \ldots \right) \\
  \PR[v_r(o)=v] = \frac1{r^2} \, \Phi\!\left(\frac{v}{r^2}\right) + \ldots
  +  \frac1{r^{\bt-1}} \, \Phi_*\!\!\left(\frac{v}{r^2}\right) + \ldots
\end{gather*}
where $\Phi(x)$ and $\Phi_*(x)$ are the inverse Laplace transforms of
$l(\xi)$ and $l(\xi)l_*(\xi)$, respectively. Probability normalization
requires $l(0)=1$ and $l_*(0)=0$; the functions $l(\xi)$ and
$l_*(\xi)$ related to $b(\xi)$ and $b_*(\xi)$ by the differential
equations:
\begin{equation*}
  \begin{split}
    \xi l'(\xi) & = - c_2 b(\xi) l(u) \\
    2 \xi l_*'(\xi) & = (\bt-3) l_*(\xi) - 2 c_2 b_*(\xi) + 
    \cs (\bt-1) b(\xi)^{\bt-2}
  \end{split}
\end{equation*}
The solution of the former is found in \cite{rndtree} and reads
\begin{equation*}
  l(\xi)  = \frac1{( \cosh \sqrt{c_2 \xi} )^2}
\end{equation*}
while for the latter we obtain
\begin{equation*}
  \begin{split}
    l_*(\xi) = \frac{\cs}{c_2} \left(\frac{\xi}{c_2}\right)^{(\bt-3)/2}
    & \left[ t^{\bt-1} 
      {}_2F_1\!\left(1,\frac{\bt-1}2;\frac{\bt+1}2;t^2 \right) \right. + \\
    & \left. + t^{\bt+1} \frac{\bt-1}{\bt+1}
      {}_2F_1\!\left(1,\frac{\bt+1}2;\frac{\bt+3}2; t^2 \right) + \right. \\
    & \left. - t^{\bt+1} \frac{\bt-2}{\bt} 
      {}_2F_1\!\left(1, \frac{\bt}2;1+\frac{\bt}2; t^2\right) \right] 
\end{split}
\end{equation*}
where $t=\tanh \sqrt{c_2 \xi}$. The solution $l_*(\xi)$ can also be
written as a power series using equation~(\ref{hyperg})
\begin{equation*}
  l_*(\xi) = \frac{\cs}{c_2} \left(\frac{\xi}{c_2}\right)^{(\bt-3)/2}
  \left[ (\bt-1) \sum_n
    \frac{t^{\bt-1+2n}}{\bt-1+2n}+ 
    \sum_n \frac{t^{\bt+1+2n}(2n+2)}{(\bt+2n)(\bt+1+2n)} \right]
\end{equation*}
from which we can extract the asymptotic behavior for vanishing $\xi$
\begin{equation*}
    l_*(\xi) \simeq \cs \xi^{\bt-2}
\end{equation*}
These results imply $\Phi(x)$ has an exponentially vanishing
$x\to\infty$ tail (see \cite{rndtree})
\begin{equation*}
  \Phi(x) \simeq \frac{\pi^2 x}{c_2^2} \, \me^{-\pi^4 x/(4 c_2)} 
\end{equation*}
while $\Phi_*(x)$ has a power-law one
\begin{equation*}
  \Phi_*(x) \simeq A\, x^{-(\bt-1)} 
\end{equation*}
Therefore we may repeat the same consideration as in the surface
probability calculation: even if $\Phi_*(x)$ is suppressed by powers
of $r$, it is the most important term for large $v/r^2$.

\subsection{$2<\bt<3$}

In this case the expansion of $g_r(\lam)$ reads
\begin{equation}
  \label{grexpB}
  g_r\left(\exp\left(-\frac{u}{r^{1/(\bt-2)}}\right)\right) 
  = 1 - \frac{a(u)}{r^{1/(\bt-2)}} + \cdots
\end{equation}
with the conditions
\begin{equation}
  \label{a_condB}
  a(0) = 0 \quad \text{and} \quad a'(0) = 1
\end{equation}
The surface probability distribution and its generating function
$G^s_r(\lam)$ can then be written as functions of scaled variables as
\begin{gather*}
  \PR[s_r(o)=s] = \frac{1}{r^{\dc-1}} \, \phi\!\left(
    \frac{s}{r^{\dc-1}} \right) + \ldots \\
  G^s_r(\exp(-u/r^{\dc-1})) = a'(u) + \ldots
\end{gather*}
where $\phi(x)$ is the inverse Laplace transform of $a'(x)$.  As
before, by substituting equation~(\ref{grexpB}) in the recurrence
equation for $g_r(\lam)$ and by equating the numerators of the
subleading terms, we obtain the differential equation
\begin{equation*}
  u a'(u) = a(u) - \cs(\bt-2)a(u)^{\bt-1}
\end{equation*}
The solution, using the conditions~(\ref{a_condB}), reads
\begin{equation}
  \label{au}
  a(u) = \frac{u}{[1+\cs(\bt-2) u^{\bt-2}]^{1/(\bt-2)}}
\end{equation}
Now, from the asymptotic behavior of $a'(u)$
\begin{equation*}
  a'(u) = \Big[1+\cs(\bt-2) u^{\bt-2}\Big]^{-(\bt-1)/(\bt-2)} \simeq 
  \begin{cases}
    1 - \cs (\bt-1) u^{\bt-2}                  & \text{for } u \to 0 \\
    [\cs(\bt-2)]^{-(\bt-1)/(\bt-2)} u^{-\bt+1} & \text{for } u \to \infty
  \end{cases}
\end{equation*}
we can determine that of the scaling function $\phi(x)$
\begin{equation*}
  \phi(x) \sim 
  \begin{cases}
    x^{-(\bt-1)} & \text{for } x \to \infty \\
    x^{\bt-2}    & \text{for } x \to 0 
  \end{cases}
\end{equation*}
which has the expected power-law behavior.

Let us now turn to the volume probability. The conditions
$h_r(1)=1$ and $h'(r)=r$ lead to the the expansion
\begin{equation*}
  h_r\left(\exp(-\xi\, r^{-\dc})\right) =
  1 - \frac{b(\xi)}{r^{\dc-1}} + \cdots 
\end{equation*}
with 
\begin{equation}
  \label{b_condB}
  b(0) = 0 \ , \qquad  b'(0) = 1
\end{equation}
The differential equation for $b(\xi)$ is found by substituting this
expansion in the recurrence (\ref{hrrec}); it reads
\begin{equation}
  \label{b_eq}
  (\bt-1)\xi b'(\xi) = b(\xi) - \cs (\bt-2) b(\xi)^{\bt-1} + (\bt-2)\xi
\end{equation}
By direct substitution one can easily verify that
\begin{equation*}
  b(\xi)=(\xi/\cs)^{1/( \bt-1)}
\end{equation*}
is a solution, but does not satisfy the second condition of
equation~(\ref{b_condB}). However we obtain a separable differential
equation if we let $b(\xi) = \xi^{1/(\bt-1)} \bb(\xi)$
\begin{equation*}
  \bb'(\xi) = \frac{\bt-2}{\bt-1}
  \left(1-\cs \bb(\xi)^{\bt-1}\right) \xi^{-\frac1{\bt-1}} 
\end{equation*}
The solution is now implicitly given by
\begin{equation*}
  \bb \; {}_2F_1 \!\! \left(1,\frac1{\bt-1};\frac{\bt}{\bt-1};\cs
    \bb^{\bt-1}\right)
  = \xi^{\frac{\bt-2}{\bt-1}}
\end{equation*}
\begin{equation*}
  \sum_{k=0}^\infty \frac1{1+k(\bt-1)} \cs^k \bb^{1+k(\bt-1)}
  = \xi^{\frac{\bt-2}{\bt-1}}
\end{equation*}
This allows us to determine the asymptotic behavior of the solution
$b(\xi)$ 
\begin{equation*}
  b(\xi) \sim
  \begin{cases}
    \xi - (\cs/\bt) \xi^{\bt-1}  & \text{for } \xi \to 0 \\
    \xi^{1/(\bt-1)}              & \text{for } \xi \to \infty
  \end{cases}
\end{equation*}

The volume probability and its scaled generating function can now be
written as
\begin{gather*}
  G^v_r(\exp(-\xi/ r^{\dc})) = l(\xi) + \ldots \\
  \PR[v_r(o)=v] = \frac1{r^{\dc}} \Phi\!\left( \frac{v}{r^{\dc}} \right)
  + \ldots
\end{gather*}
where $l(\xi)$ is the Laplace transform of $\Phi(x)$. As before
$l(0)=1$ by normalization and $l(\xi)$ is obtained from the
differential equation
\begin{equation}\label{heq}
  \xi l'(\xi) = - \cs(\bt-2) l(\xi) b(\xi)^{\bt-2}
\end{equation}

The small and large $\xi$ asymptotic behavior of $l(\xi)$ can be
obtained from those of $b(\xi)$, and read 
\begin{equation}\label{hasym}
  l(\xi) \simeq 
  \begin{cases}
    1 - \cs \xi^{\bt-2}                    & \text{for } \xi \to 0 \\
    \exp(-\cs(\bt-1)\xi^{(\bt-2)/(\bt-1)}) & \text{for } \xi \to \infty
  \end{cases}
\end{equation}
Therefore the asymptotic behavior of $\Phi(x)$ reads
\begin{equation*}
  \Phi(x) \sim
  \begin{cases}
    x^{-(\bt-1)}                 & \text{for } x \to \infty \\
    x^{-\bt/2} \exp(-k x^{2-\bt}) & \text{for } x \to 0
  \end{cases}
\end{equation*}
with
\begin{equation*}
  k = \cs^{\bt-1} (\bt-2)^{\bt-2}
\end{equation*}
The probability distribution is exponentially vanishing for small
volumes and it has a power law tail with exponent $1-\bt$, as expected. 

% \section{Moments of volume: $\expv{v_r^k}_V$}
% \label{sec:moments}

% For higher moments the only difference stands in the expansion of
% $\Big[1+\sum_j v_j\Big]^n$; the $n=3$ case gives
% \begin{equation*}
%   \begin{split}
%     \Eb{3}(r,\lam) = &  \lam g(\Gb(\lam)) + \\
%     & \lam g'(\Gb(\lam))\Big[ 3 \Eb{1}(r-1,\lam) 
%     + 3 \Eb{2}(r-1,\lam) + \Eb{3}(r-1,\lam) \Big] + \\
%     & \lam g''(\Gb(\lam))\Big[ 3 \Eb{1}(r-1,\lam)^2 
%     + 3 \Eb{1}(r-1,\lam) \Eb{2}(r-1,\lam)\Big] + \\
%     & \lam g'''(\Gb(\lam))  \Eb{1}(r-1,\lam)^3
%   \end{split}
% \end{equation*}
% This can be easily generalized to every $n\ge3$: for every $k\le n$
% the $k$-th derivative $g^{(k)}(\Gb(\lam))$ multiplies a product of
% $k$ expectation values.
% Now we can prove by induction that
% \begin{equation}
%   \label{evn}
%   \Eb{n}(r,\lam) \simeq r^{n+1} (1-\lam)^{(\bt-1-n)/(\bt-1)}
% \end{equation}
% Indeed the $n=1$ and $n=2$ cases are already verified; for
% $n\ge3$ one can easily see that the only relevant terms are
% \begin{equation*}
%    \Eb{n}(r,\lam) - \lam g'(\Gb(\lam))  \Eb{n}(r-1,\lam)
%    \simeq \lam g^{(n)}(\Gb(\lam)) \Eb{1}(r-1,\lam)^n
% \end{equation*}
% because 
% \begin{equation*}
%   g^{(n)}(\Gb(\lam)) \simeq (1-\lam)^{(\bt-1-n)/(\bt-1)} \qquad
%   \text{for } n\ge2
% \end{equation*}
% Equation~(\ref{evn}) is then easily recovered, yielding
% equation~(\ref{expvn}).

\section{$\V{n}$ moments}
\label{sec:vn-averages}
First of all let us write the explicit form of
$\cF_n(\V1,\V2,\ldots,\V{n-1})$
\begin{equation*}
  \cF_n(\V1,\V2,\ldots,\V{n-1}) = 
  \sum_{k=2}^n \sum_{n_1=1}^{n-1}\ldots\sum_{n_k=1}^{n-1}
  \delta\Big(n-\sum_{j=1}^k n_i\Big) \prod_{j=1}^k \V{n_j}
\end{equation*}
Now we can write an expression for $\Ebv{\V{n}}$ and its
generating function $\Ebl{\V{n}}$ involving only
$\V1,\V2,\ldots,\V{n-1}$:
\begin{equation*}
  \begin{split}
    \Ebv{\V{n}} & 
    = \sum_z f_z \sum_{\{\VV{k}j\}} \dl\Big(\V1-1-\sum_j \VV1j\Big)
    \sum_j \Big\{\VV{n}j+\cF_n(\VV1j,\VV2j,\ldots,\VV{n-1}j)\Big\} \prod_j 
    \Pb(\{\VV{k}j\}) \\
    & = \sum_z f_z \sum_{\VV1j} \dl\Big(\V1-1-\sum_j \VV1j\Big) 
    \sum_j \Big\{\Ebv[\VV1j]{\V{n}}+
    \Ebv[\VV1j]{\cF_n(\V1,\V2,\ldots,\V{n-1})} \Big\} 
    \prod_{j'\neq j} \Pb(\VV1{j'})
  \end{split}
\end{equation*}
\begin{equation*}
  \begin{split}
    \Ebl{\V{n}}  & = \lam \sum_z f_z (z-1) \Big\{ \Ebl{\V{n}} + 
    \Ebl{\cF_n(\V1,\V2,\ldots,\V{n-1})} \Big\} \Gb(\lam)^{z-2} \\ 
    & = \lam \, g'(\Gb(\lam)) 
    \Big\{ \Ebl{\V{n}} + \Ebl{\cF_n(\V1,\V2,\ldots,\V{n-1})} \Big\} \\
    & = \Big[1-\lam \, g'(\Gb(\lam))\Big]^{-1} \lam \, g'(\Gb(\lam))
    \Ebl{\cF_n(\V1,\V2,\ldots,\V{n-1})}
  \end{split}
\end{equation*}
This shows that $\Ebl{\V{n}}$ is more singular than
$\Ebl{\cF_n(\V1,\V2,\ldots,\V{n-1})}$, which has now to be calculated.
To do this we need an expression for the generic multiple moment
$\Ebl{\V1^{k_1}\V2^{k_2}\cdots\V{n}^{k_n}}$. First of all notice that
powers of $\V1$ can be extracted, that is
\begin{equation*}
  \Ebv{\V1^{k_1}\V2^{k_2}\cdots\V{n}^{k_n}} = 
  V^{k_1} \Ebv{\V2^{k_2}\cdots\V{n}^{k_n}}
\end{equation*}
The recurrence rule for $\Ebl{\V2^{k_2}\cdots\V{n}^{k_n}}$ reads
\begin{equation}
  \label{genrec}
  \begin{split}
    \Ebl{\V2^{k_2}\cdots\V{n}^{k_n}} = 
    \Big[1-\lam \, g'(\Gb(\lam))\Big]^{-1}
    \Big\{ & \lam \, g'(\Gb(\lam)) \sum_{j=2}^n 
    \Ebl{\V2^{k_2}\ldots k_j\V{j}^{k_j-1} \cF_j \cdots \V{n}^{k_n}} + \\
    & \lam \, g''(\Gb(\lam)) \sum_{k_{j,1},k_{j,2}} 
    \Ebl{\V2^{k_{2,1}}\cdots\V{n}^{k_{n,1}}}
    \Ebl{\V2^{k_{2,2}}\cdots\V{n}^{k_{n,2}}} + \\
    & \cdots \\
    & \lam \, g^{(N)}(\Gb(\lam)) \sum_{k_{j,1},\ldots,k_{j,N}} 
    \prod_{l=1}^N \Ebl{\V2^{k_{2,l}}\cdots\V{n}^{k_{n,l}}} 
    \Big\}
  \end{split} 
\end{equation}
where $N=\sum_j k_j$, the sums over $k_{j,l}$ are such that $\sum_l
k_{j,l}=k_j$, and only most singular terms are kept.
Equation~(\ref{genrec}) can be recursively used; the moments on the 
right-hand side either involve products with fewer terms or lower $j$ $\V{j}$'s
so that we end with moments of $\V1$, which are trivial. 

The general expression of $\Ebl{\V2^{k_2}\cdots\V{n}^{k_n}}$ for
both $2<\bt<3$ and $\bt>3$ cases 
\begin{equation}
  \label{fbres}
  \Ebl{\V1^{k_1}\V2^{k_2}\cdots\V{n}^{k_n}} \simeq
  \begin{cases}
  (1-\lam)^{[1 - \sum_j k_j ((2\bt-3)j -\bt+2)]/(\bt-1)} &
  2<\bt<3 \\
  (1-\lam)^{\frac12 - \frac12 \sum_j k_j (3j-1)} & \bt>3
  \end{cases}
\end{equation}
can now be proved by induction. Substituting equation~(\ref{fbres}) in
the right-hand side of equation~(\ref{genrec}) we see that all terms
have the same singular behavior when $2<\bt<3$, while the
$g'(\Gb(\lam))$ and $g''(\Gb(\lam))$ terms are the leading order ones
for $\bt>3$.

The branch probability moments just calculated are related to the tree
probability ones by
\begin{equation*}
  \begin{split}
    \Etl{\V{n}} & = \lam \, \ghat'(\Gb(\lam)) 
    \Big[ \Ebl{\V{n}} + \Ebl{\cF_n(\V1,\V2,\ldots,\V{n-1})} \Big]  \\
    & = \lam \, \ghat'(\Gb(\lam)) \Ebl{\V{n}} + \ldots 
  \end{split}
\end{equation*}
and
\begin{equation*}
  \Etl{\V2^{k_2}\cdots\V{n}^{k_n}} = \lam \, \ghat'(\Gb(\lam))
  \Ebl{\V2^{k_2}\cdots\V{n}^{k_n}} + \ldots
\end{equation*}
Therefore $\Etl{\V1^{k_1}\cdots\V{n}^{k_n}}$ has the same asymptotic
behavior (equation~(\ref{fbres}) as
$\Ebl{\V1^{k_1}\cdots\V{n}^{k_n}}$. This imply the following
asymptotic behavior of
$\Etv{\V1^{k_1}\V2^{k_2}\cdots\V{n}^{k_n}}$ 
\begin{equation*}
  \Etv{\V1^{k_1}\V2^{k_2}\cdots\V{n}^{k_n}} \simeq
  \begin{cases}
    V^{[\sum_j k_j ((2\bt-3) j -\bt+2) - \bt]/(\bt-1)}
    & 2<\bt<3 \\
    V^{\frac12 \sum_j k_j (3 j - 1) - \frac32} & \bt>3
  \end{cases}
\end{equation*}
and finally 
\begin{equation} \label{expvVn}
  \expv{\V1^{k_1} \V2^{k_2} \cdots \V{n}^{k_n}}_V \simeq
  \begin{cases}
    V^{\sum_j k_j ((2\bt-3) j -\bt+2)/(\bt-1)}
    & 2<\bt<3 \\
    V^{\frac12 \sum_j k_j (3 j - 1)} & \bt>3
  \end{cases}
\end{equation}


\begin{thebibliography}{99}
\bibitem{nab} M. E. J. Newman, preprint cond--mat/0202208; \\
  R. Albert, A.-L. Barabasi, {\em Rev. Mod. Phys.} {\bf 74}, 47
  (2002).
\bibitem{percol} D. Stauffer and A. Aharony, {\em Introduction to
    percolation theory}, Taylor \& Francis, London (1994).
\bibitem{branched} J. Ambjorn, B. Durhuus and T. Jonsson, \emph{Phys.\ 
    Lett.\ B } {\bf 244}, 403 (1990);\\
  J. Ambjorn, B. Durhuus and T. Jonsson, {\em Quantum Geometry},
  Cambridge, 1997.
\bibitem{bollobas}  B. Bollobas, \emph{Random Graphs}, Academic Press,
  New York (1985). 
\bibitem{barabasi} A.-L. Barabasi, R. Albert \emph{Science}
  {\bf 286}, 509 (1999).
\bibitem{exotic} Z. Burda, J. Erdmann, B. Petersson, M. Wattenberg,
  {\em Phys. Rev. E} {\bf 67}, 026105 (2003).
\bibitem{bck} Z. Burda, J. D. Correia, A. Krzywicki {\em Phys. Rev. E}
  {\bf 64}, 046118 (2001).
\bibitem{shlomo} S. Havlin and D. Ben--Avraham, {\em Adv. Phys.}
{\bf 36} 695--798 (1987).
\bibitem{rndtree}  C. Destri and L. Donetti, {\em J. Phys. A} {\bf
    35}, 5147 (2002).
\bibitem{suzuki} 
  M. Suzuki, \emph{Progr.\ Theor.\ Phys.} {\bf 69}, 65 (1983)
\bibitem{alorb} S. Alexander and R. Orbach, {\em J. Phys. Lett.}  {\bf
    43}, 625 (1982).
\bibitem{combntd} R. Burioni, D. Cassi, {\em Phys. Rev. E} {\bf 51},
  2865 (1995) \\
  D. Cassi and S. Regina, \emph{Mod.\ Phys.\ Lett. B} {\bf 6}, 1397
  (1992).
\bibitem{hhw} K. Hattori, T. Hattori and H. Watanabe, \emph{Progr.\ 
    Theor.\ Phys.\ Suppl.} {\bf 92}, 108 (1987).
\bibitem{gwproc} K.B. Athreya, P.E. Ney, {\em Branching Processes},
  Springer-Verlag, New York (1972).
\bibitem{kesten} H. Kesten, {\em Ann.\ Inst.\ H. Poincar\'e Probab.\ 
    Statist.} {\bf 22}, 425 (1987).
\bibitem{rndtree2} C. Destri and L. Donetti, {\em J. Phys. A} {\bf
    35}, 9499 (2002).
\end{thebibliography}
\end{document}